\documentclass[twoside]{mhd}

 \usepackage{graphicx} 
\usepackage{authblk}
 \usepackage{amsmath}
\usepackage{fancyhdr}
\usepackage{xcolor}
\usepackage{subcaption}
\usepackage{hyperref}
\usepackage{soul}
\title{Modeling the free-surface magnetohydrodynamics of thick liquid metal walls for fusion }

\author{Valentina Giovacchini, Eric Favre and Francesco A. Volpe}
\institute{Renaissance Fusion, 14 rue J. P. Timbaud, 38600 Fontaine, France\\
 \small $^*$valentina.giovacchini@renfusion.eu \\
 Preprint submitted to \textit{Magnetohydrodynamics}, 29/11/2024
}

\begin{document}
\maketitle

\begin{abstract}
Renaissance Fusion proposes thick liquid metal walls as plasma-facing components for future commercial fusion reactors. It designs and operates proof-of-concept experiments aiming at actively suspending and stabilizing a flowing free-surface liquid metal layer against gravity using Lorentz forces. The first operating prototype consists of a $1 \, \text{m}$-diameter chamber in the presence of a magnetic field of the order of $0.3 \, \text{T}$. A GaInSn flow is injected inside the chamber, and it is actively suspended thanks to the injection of currents up to $3 \, \text{kA}$. To simulate the prototype, a numerical tool capable of modeling magnetohydrodynamics phenomena in two-phase flows has been developed to reproduce and interpret the experimental results. The tool relies on the low-magnetic Reynolds assumption, implements the Volume-of-Fluid method for tracking the free-surface, and couples the electric potential in both fluid and solid domains.
\end{abstract}

\section{Introduction}
A promising design for commercial fusion reactors incorporates flowing liquid metals as plasma-facing material, simultaneously serving to extract heat and absorb neutrons. This approach provides a continuously regenerated surface, thereby extending the lifetime of the underlying solid substrate components \cite{tabares2015present}. A liquid metal composition including lithium allows for the direct breeding of tritium in-vessel, followed by its subsequent harvesting downstream. Furthermore, using liquid lithium as a plasma-facing material has been associated with enhanced plasma confinement performance, as it enables low recycling regimes \cite{de_castro_lithium_2021}.

By leveraging the intense plasma-confining magnetic field, it is possible to actively suspend thick layers of liquid metal by carefully balancing centrifugal and Lorentz forces through the injection of electrical currents \cite{abdou2001exploration, morley2004progress, mirhoseini2017passive}. Renaissance Fusion is designing and operating proof-of-concept experiments to demonstrate this gravity-defying concept. The first operating prototype, called Skyfall 1b, consists of a $1 \, \text{m}$-diameter, $3 \, \text{cm}$-wide chamber with a magnetic field of approximately $0.3 \, \text{T}$ generated by permanent magnets. A GaInSn flow is forced inside the chamber and suspended by injecting currents up to $3 \, \text{kA}$.

Simulating Skyfall 1b requires a specialized numerical tool capable of capturing the complex magnetohydrodynamic (MHD) behavior in two-phase flows. Within the OpenFOAM-v2212 framework, we developed a model that computes the magnetic field in a pre-processing phase, in the limit of low magnetic Reynolds number $R_m$ \cite{davidson2017introduction} such that the magnetic field remains unaffected by the fluid dynamics. 
This limit is justified because, as will be shown in Section \ref{sec:model},
$R_m$ is calculated to be 0.13 for the experiment of interest, while the applicability of this assumption to future experiments is discussed in the conclusions.
A conservation equation for the electric potential models the interactions between the current density and the fluid flow. Additionally, the injection of external currents through electrodes generates controllable Lorentz forces, enabling manipulation of the flow configuration alongside gravitational forces.

To accurately track the free-surface dynamics, we employ a Volume-of-Fluid (VOF) method \cite{deshpande2012evaluating}, which captures the interface between the liquid and gas phases, while the Navier-Stokes equations describe both phases, with the liquid treated as an incompressible, Newtonian, isothermal flow. The uniform temperature assumption is valid for this experiment due to minimal heat loads but would not apply in reactor conditions with significant temperature gradients. 
To model turbulence, $k$-$\omega$ Shear Stress Transport (SST) \cite{menter1994two} model is applied within the fluid domain, where $k$ is the turbulent kinetic energy and $\omega$ its specific dissipation rate.

A series of experiments at varying flow rates and current levels provides data on free-surface velocity, jet thickness, and landing point position,  essential for model validation \cite{EXPERIMENTALPAPER}. The numerical tool predicts magnetic field distribution, jet trajectory, thickness, and velocity, with results benchmarked against these experimental measurements.

Section \ref{sec:model} details the model assumptions and equations implemented, followed by numerical results and a critical analysis of the findings in Section \ref{sec:results}, highlighting both the achievements and areas for further model refinement in Section \ref{sec:conclusion}.

\section{Numerical model and governing equations}\label{sec:model}
\begin{figure}[htb!]
\centering
\vspace{-0.2cm}
 \begin{subfigure}{0.4\textwidth}
        \centering \caption{ \label{fig:sketch} }
        \includegraphics[height=0.27\textheight]{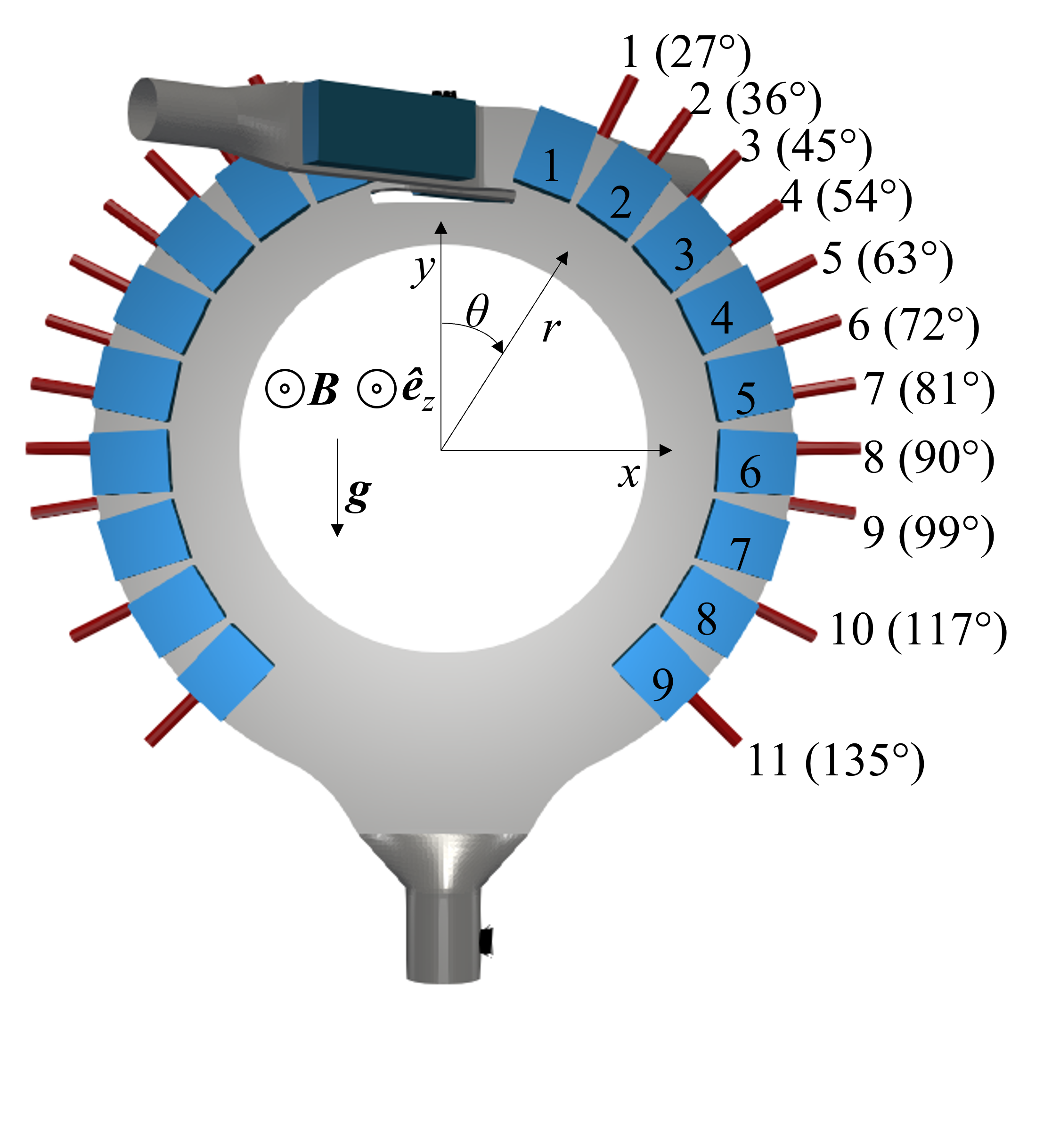}
    \end{subfigure}\hspace{0.2cm}
     \begin{subfigure}{0.55\textwidth}
        \centering \caption{ \label{fig:BC} }
        \includegraphics[height=0.27\textheight]{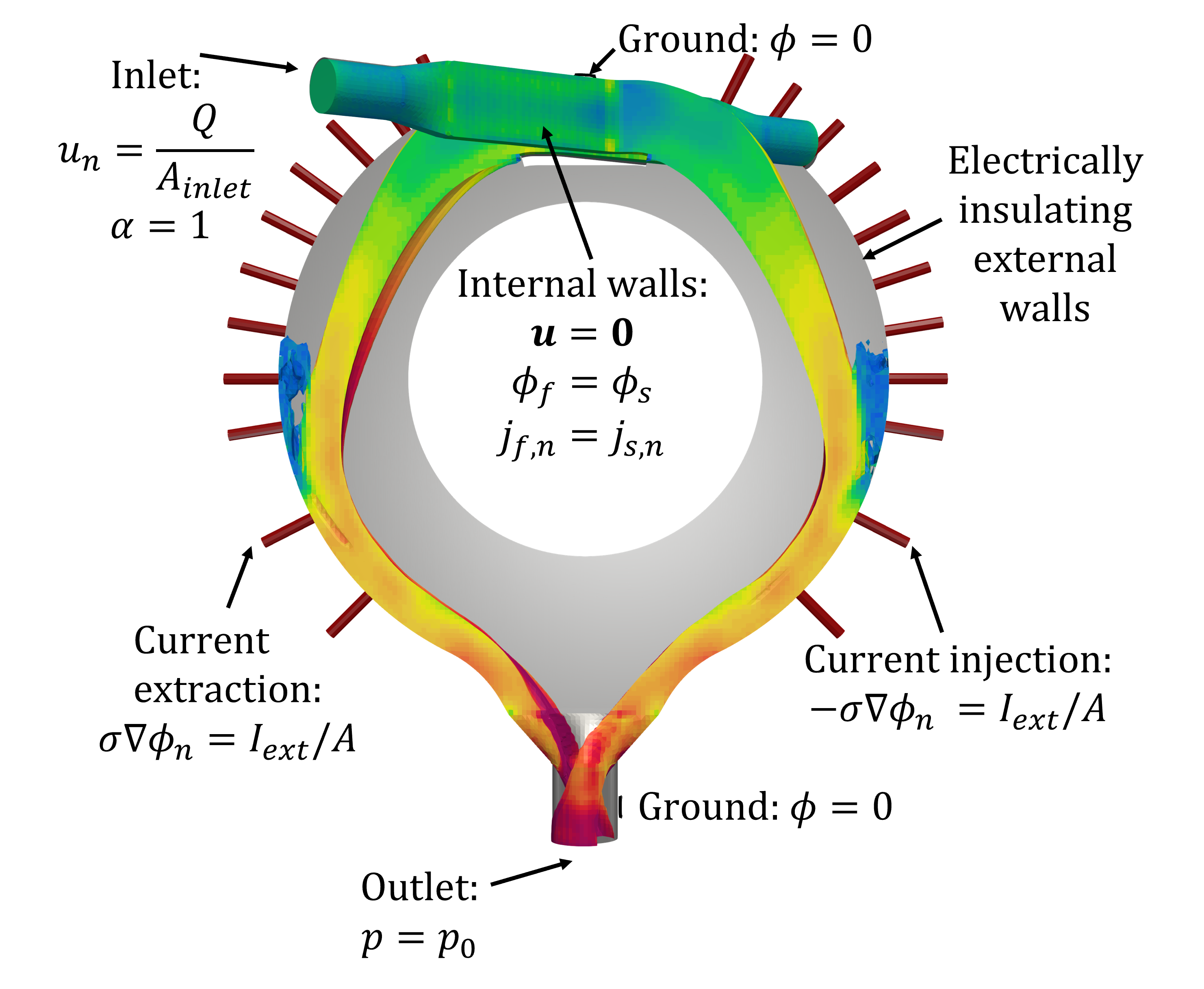}
    \end{subfigure}
\caption{Numerical model of the Skyfall 1b vacuum chamber: (a) illustrates the magnets (1-9), electrodes (1-11), and steel components, along with the Cartesian $(x, y)$ and polar $(r, \theta)$ coordinate systems; (b) depicts a representative liquid metal domain and the applied boundary conditions, with $A$ denoting the cross section of the electrodes.}
\vspace{-0.5cm}
\end{figure}
\subsection{Experimental setup}
The Skyfall 1b experiment consists of a GaInSn loop connected to a vacuum chamber with a diameter of $1 \, \text{m}$ and a width of $3 \, \text{cm}$ along $z$. The fluid is set in motion by a permanent magnet pump, providing flow rates of up to $11 \, \text{L/s}$. The fluid is pumped into the top of the chamber and sucked out from the bottom. Two rings of permanent magnets generate inside the chamber a nonuniform magnetic field $\mathbf{B}$ in the $z$-direction, with a maximum magnitude of approximately $0.3 \, \text{T}$. Each side wall is equipped with 11 electrodes. Power supplies inject currents of up to $3 \, \text{kA}$. More details are available in \cite{EXPERIMENTALPAPER}. The geometrical model developed in OpenFOAM is illustrated in Figure \ref{fig:sketch}. The magnet domain is shown in blue, stainless steel in gray, electrodes in red.

\subsection{Low magnetic Reynolds number and field perturbations} The physical model is based on the low magnetic Reynolds number \( R_m \) assumption,
which implies that the magnetic field generated by the induced current, \( \mathbf{J} \sim \sigma (\mathbf{u} \times \mathbf{B}) \),  where $\sigma $ and $\mathbf{u}$ are the electrical conductivity and the velocity field respectively,
is negligible compared to the externally imposed magnetic field. According to Ampère’s law, the magnitude of the magnetic field
associated with the induced current is \( |\mathbf{b}| \sim \mu l \sigma u B \sim R_m B \), where $\mu$ and $l$ are the magnetic permeability and the characteristic length respectively. Therefore, when \( R_m \) is small, 
\( |\mathbf{b}| \ll B \), justifying the approximation \cite{davidson2017introduction}.
 Considering a 0.1 m thick layer of flowing GaInSn with a velocity of 1 m/s, electrical conductivity of  \( 3.4 \times 10^6 \, \text{S/m} \), relative permeability equal to 1 and magnetic permeability $\mu=4 \pi\ 10^{-7}$ H/m, the magnetic Reynolds number is \( R_m =\mu l \sigma u= 0.13 \). 
 
 
 Another source of magnetic field perturbation is the injection of external current through the electrodes. If we consider an injection of 1 kA through the liquid metal layer with a cross-sectional area of \( 0.003 \, \text{m}^2 \), the resulting magnetic field is \( |\mathbf{b}| \sim l \mu J \approx 0.013 \, \text{T} \), which is roughly the 5\% of the imposed magnetic field, and can be neglected as first approximation. 


\subsection{Magnetic field}\label{sec:magneticField}
The solver \textit{magneticFoam} is used to compute the magnetic field map generated by the permanent magnets in the experiment. In the current-free region, $\nabla \times \mathbf{H} = 0$, which implies that $\mathbf{H}$ can be expressed as the gradient of a scalar potential, i.e., $\mathbf{H} = -\nabla \psi$, where $\psi$ is the magnetic scalar potential. From Maxwell's equation, $\nabla \cdot \mathbf{B} = \mu_0 \mu_r \nabla \cdot (\mathbf{H} + \mathbf{M})$, we obtain the governing equation
\begin{equation}
    \nabla \cdot (\mu_0 \mu_r \nabla \psi) = \nabla \cdot \mathbf{M}\,,
\end{equation}
where $\mathbf{M}$ represents the magnetization of the permanent magnets,  $\mu_0$ is the permeability of vacuum, and $\mu_r$ is the relative permeability of the medium. The magnetic field can then be evaluated as
\begin{equation}\label{eq:B}
    \mathbf{B} = -\mu_0 \mu_r \nabla \psi + \mathbf{M}.
\end{equation}
\subsection{Multiregion MHD two-phase flow model}\label{sec:MHD}
In this section, we present the in-house multiregion model developed in OpenFOAM v2212. To simulate the conducting walls and their influence on flow magnetohydrodynamics, the fluid domain is coupled to the solid domain through the electric potential $\phi$. The governing equations are iteratively solved in the different domains, within a PIMPLE loop \cite{holzmann2016mathematics}, until convergence is reached for each time step.

\noindent
\subsubsection{{Fluid region}}
The liquid GaInSn at room temperature is forced into the vacuum chamber shown in Figure \ref{fig:sketch}, where argon atmosphere is maintained at approximately $0.1 \, \text{mbar}$. To model the free-surface flow, we rely on the VOF method, which models GaInSn as liquid and argon as  gas. This method is implemented as the \textit{interFoam} solver \cite{ubbink1997numerical}. 

Both fluids are considered incompressible, isothermal, and immiscible. A single set of conservation equations is expressed for the mixture, with the properties of the fluid computed as a weighted average based on the volume fraction $\alpha$. Specifically, a generic property $\varphi$ is evaluated as
$    \varphi = \varphi_1 \alpha + \varphi_2 (1 - \alpha)\,,$
where $\alpha = 1$ indicates that the cell is entirely occupied by the liquid phase, and $\alpha = 0$ indicates the cell is occupied entirely by the gas phase. Cells with $0 < \alpha < 1$ represent the interface. The volume fraction satisfies the following transport equation:
\begin{equation}\label{eq:alpha}
    \frac{\partial \alpha}{\partial t} + \nabla \cdot (\mathbf{u} \alpha) = 0\,.
\end{equation}
In \textit{interFoam}, the advection term is modified to ensure that the transport of $\alpha$ preserves its sharpness and reduces numerical diffusion \cite{deshpande2012evaluating}.

Starting from the conservation equations of mass and momentum implemented in \textit{interFoam}, the solver has been modified to include magnetohydrodynamic effects. The governing equations for the mixture are given by:
\begin{equation}
\begin{aligned}\label{eq:NS}
    \nabla \cdot \mathbf{u} &= 0 \\
    \frac{\partial \rho \mathbf{u}}{\partial t} + \nabla \cdot (\rho \mathbf{u} \otimes \mathbf{u}) &= - \nabla p + \rho \mathbf{g} + \nabla \cdot \mathbf{\tau}_{\text{eff}} + \mathbf{f}_\gamma + \mathbf{f}_L\,,
\end{aligned}
\end{equation}
where $p$ is the pressure field, $\rho$ is the density of the mixture, $\mathbf{g}$ is the gravity vector, 
$\mathbf{\tau}_{\text{eff}} = \mathbf{\tau} + \mathbf{\tau}_R$ is the effective stress tensor, which includes both the viscous stress tensor $\mathbf{\tau}$ and the Reynolds stress tensor $\mathbf{\tau}_R$. The viscous stress tensor $\mathbf{\tau}$ is given by $\mu (\nabla \mathbf{u} + \nabla \mathbf{u}^\top)$ for Newtonian fluids, where $\mu$ is the dynamic viscosity of the mixture. The Reynolds stress tensor $\mathbf{\tau}_R$ is modeled as $\mu_T (\nabla \mathbf{u} + \nabla \mathbf{u}^\top)$, with a $k$-$\omega$ SST model to evaluate the eddy viscosity $\mu_T$. The surface tension force per unit volume $\mathbf{f}_\gamma$ is described using a Continuum Surface Force (CSF) model \cite{brackbill1992continuum}, which assumes $\mathbf{f}_\gamma = \gamma \kappa \nabla \alpha$, where $\gamma$ is the surface tension and $\kappa$ is the curvature of the interface. 

The Lorentz force per unit volume $  \mathbf{f}_L = \mathbf{J} \times \mathbf{B}$ has been included following the procedure described in \cite{ni2007current2}. The current density $\mathbf{J}$ is given by Ohm's law, $\mathbf{J} = \sigma (- \nabla \phi + \mathbf{u} \times \mathbf{B})$, with $\phi$ defined by $\mathbf{E} = -\nabla \phi$, while the magnetic field $\mathbf{B}$ is evaluated as reported in Equation \eqref{eq:B}.
The principle of conservation of electric charge gives $\nabla \cdot \mathbf{J} = 0$, and the electric potential satisfies the Poisson equation:
\begin{equation}\label{eq:potFluid}
    \nabla \cdot (\sigma \nabla \phi) = \nabla \cdot (\sigma \mathbf{u} \times \mathbf{B})\,.
\end{equation}

\noindent
\subsubsection{{Solid region}}
The stainless-steel structure and the copper rods visible in Figure \ref{fig:sketch} are modeled as rigid, isothermal solid regions. Here, Ohm's law reduces to $\mathbf{J} = -\sigma \nabla \phi$. Thus, the electric potential satisfies the Laplace equation $
    \nabla \cdot (\sigma \nabla \phi) = 0$.

\noindent
\subsubsection{{Boundary conditions}}
The boundary conditions are reported in Figure \ref{fig:BC}. The inlet section is modeled as a Dirichlet boundary condition for the velocity field, while at the outlet, we impose the total pressure field $p_0$. These surfaces are considered electrically insulating, meaning $\mathbf{J} \cdot \mathbf{n} = 0$, which translates to $-\sigma \nabla \phi \cdot \mathbf{n} = -\sigma \mathbf{u} \times \mathbf{B} \cdot \mathbf{n}$.
The no-slip walls represent the interface between the fluid region and the solid domains. The interface satisfies the following coupling conditions:
\begin{equation}
\label{eq:coupling}
    \phi_i = \phi_j\,, \qquad
    -\sigma_i \nabla \phi_i \cdot \mathbf{n} = -\sigma_j \nabla \phi_j \cdot \mathbf{n} \qquad \text{at the interface}\,,
\end{equation}
where $\mathbf{n}$ represents the vector normal to the surface of the interface between region $i$ and region $j$. 

The ground in the solid domain is modeled as a homogeneous Dirichlet boundary condition $\phi = 0$. The external walls in contact with the surrounding air are modeled as electrically insulating, which leads to $-\sigma \nabla \phi \cdot \mathbf{n} = 0$. The outer surfaces of the copper rods connected to the power supply are modeled with inhomogeneous Neumann boundary conditions satisfying the relation  $\mathbf{J} \cdot \mathbf{n} = -\sigma \nabla \phi \cdot \mathbf{n} = {I_{ext}}/{A}\,,$ 
where $A$ is the cross-section of the copper rod and $I_{ext}$ is the external current injected.

\section{Numerical Results}\label{sec:results}
\subsection{Magnetic field}
\begin{figure}[htb!]\vspace{-0.5cm}
    \centering
    \begin{subfigure}{0.45\textwidth}
        \centering \caption{ \label{fig:B(r)} }
        \includegraphics[height=0.76\textwidth]{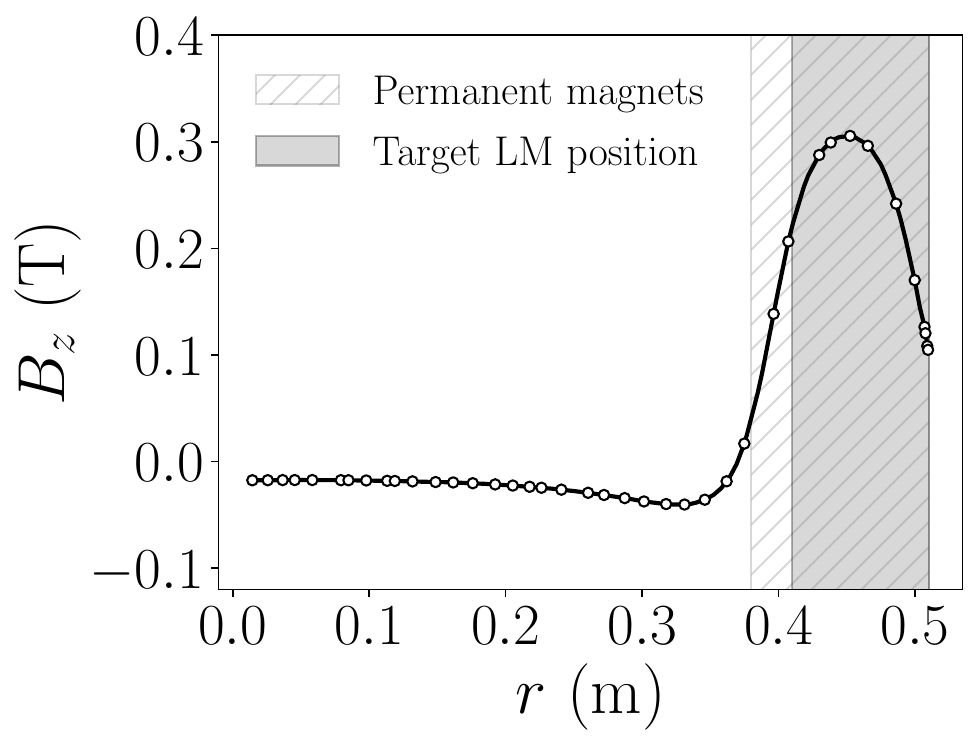}
    \end{subfigure}
    \begin{subfigure}{0.45\textwidth}
        \centering        \caption{}
        \label{fig:B(theta)}
        \includegraphics[height=0.76\textwidth]{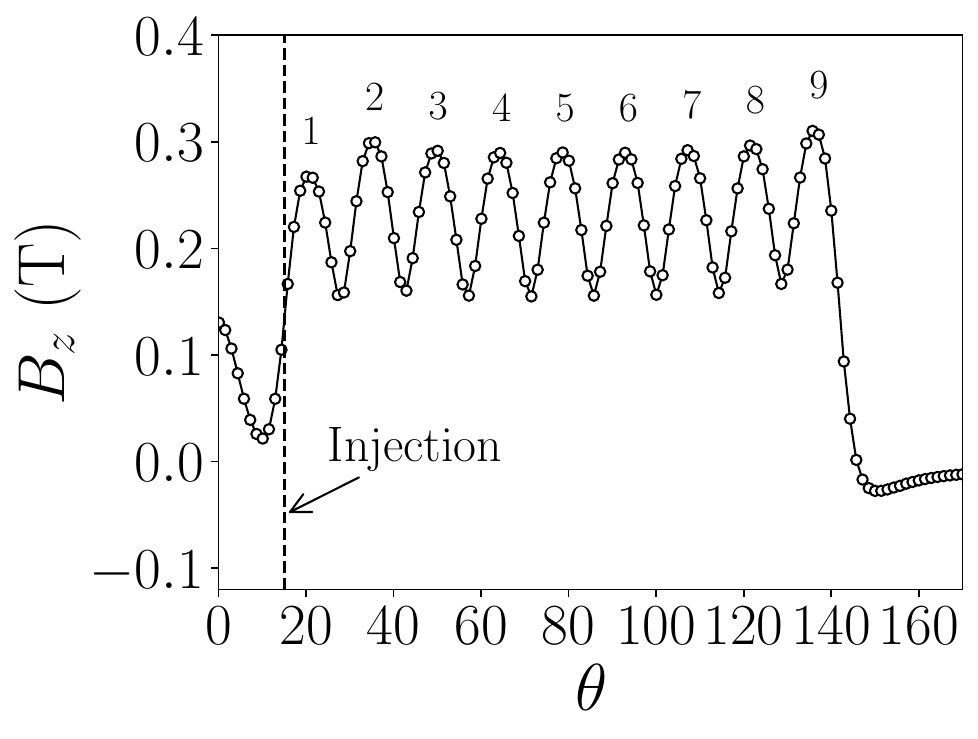}
    \end{subfigure}
     \vspace{-0.5cm}
    \caption{Plots of the magnetic field $z$-component at $z = 0 \, \text{m}$: (a) as a function of the radial position $r$ at \(\theta = 135^\circ\), and (b) as a function of the angular position $\theta$ at \(r = 0.45 \, \text{m}\). In (a), the radial position occupied by the permanent magnets and the target liquid metal (LM) position are highlighted. In (b), the magnets (numbered 1 to 9, as defined in Figure \ref{fig:sketch}) and the injection position at $\theta = 15^\circ$ are indicated.}
     \vspace{-0.5cm}
\end{figure}
The magnetic field distribution is evaluated following the methodology outlined in Section \ref{sec:magneticField}. A cubic computational domain with a side length of \(2 \, \text{m}\) is discretized into 11.6 million hexahedral cells, with mesh refinement applied near the chamber and the magnets. Each magnet has dimensions of \(19.5 \, \text{mm} \times 89 \, \text{mm} \times 110.6 \, \text{mm}\), and a magnetization $|\mathbf M|$ of \(1.03 \, \text{MA/m}\) is assigned. The relative magnetic permeability $\mu_r$ is set to 1 throughout the domain.

In Figure \ref{fig:B(r)}, we show the $z$-component of the magnetic field, \(B_z\), as a function of the radial coordinate $r$ for a fixed angular position \(\theta = 135^\circ\), at \(z = 0 \, \text{m}\). The longitudinal component reaches its maximum 0.31 T around $r=0.45$ m.
Figure \ref{fig:B(theta)} illustrates the variation of \(B_z\) as a function of the angular position \(\theta\), for a fixed radius of \(r = 0.45 \, \text{m}\). The field exhibits substantial non-uniformity along the arc, oscillating between 0.15 T and 0.31 T. The maximum longitudinal field is obtained in correspondence of the nine pairs of permanent magnets.
\subsection{Fluid flow}
\begin{table}[htb!]\centering
\caption{\label{tab:nondimensional} Reynolds $Re$, Hartmann $Ha$, interaction $N$, Froude $Fr$ and Weber $Wb$ numbers evaluated for  different flow rates $Q$.}
\vspace{-0.5cm}
\[\begin{array}{|c|c|c|c|c|c|}\hline
&Re	&   Ha & N	    &	Fr    & Wb         \\\hline
Q \text{ (L/s)} & \frac{\text{inertia}}{\text{viscous}} & \frac{\text{lorentz}}{\text{viscous}} & \frac{\text{lorentz}}{\text{inertia}} & \frac{\text{inertia}}{\text{gravity}} & \frac{\text{inertia}}{\text{surface tension}}\\\hline
4.0 &9.9 \ 10^{4}	& 3.4 \ 10^2&	1.20	&	0.38  & 1.5 \ 10^{2}
                       \\
5.9 &1.5 \ 10^{5}	& 3.4 \ 10^2&	0.78	&	0.56   & 3.3 \ 10^{2}                     \\
9.2 &2.3\ 10^{5}	& 3.4 \ 10^2&	0.50	&	0.87   & 8.0 \ 10^{2}                   \\
10.5 &2.6	\ 10^{5}& 3.4 \ 10^2&	0.44	&	1.00 &1.0 \ 10^{3}
\\\hline
\end{array}\]
 \vspace{-0.5cm}
\end{table}
The chamber model, depicted in Figure \ref{fig:sketch}, is discretized into 560,000 hexahedral cells.
The mesh employs a maximum cell size of 8 mm, with 8 cells in the boundary layers and a thickness ratio of 1.2. The fluid, steel and copper regions are solved iteratively at each time step as described in Section \ref{sec:MHD}.
The numerical results presented in the following refer to a range of flow rates: \(4.0\), \(5.9\), \(9.2\), and \(10.5 \, \text{L/s}\), reflecting specific experimental settings of the electric motor. The associated non-dimensional numbers, reported in Table \ref{tab:nondimensional}, indicate that viscous forces and surface tension are negligible compared to inertia, gravity, and Lorentz forces.

\noindent
\subsubsection{{Base  flow - no external current}}
\begin{figure}[htb!]\vspace{-0.5cm}
\centering
\begin{subfigure}{0.45\textwidth}
\caption{$Q = 5.9$ L/s, $I = 0$ A\label{fig:5.9L/s}}
\includegraphics[width =\textwidth]{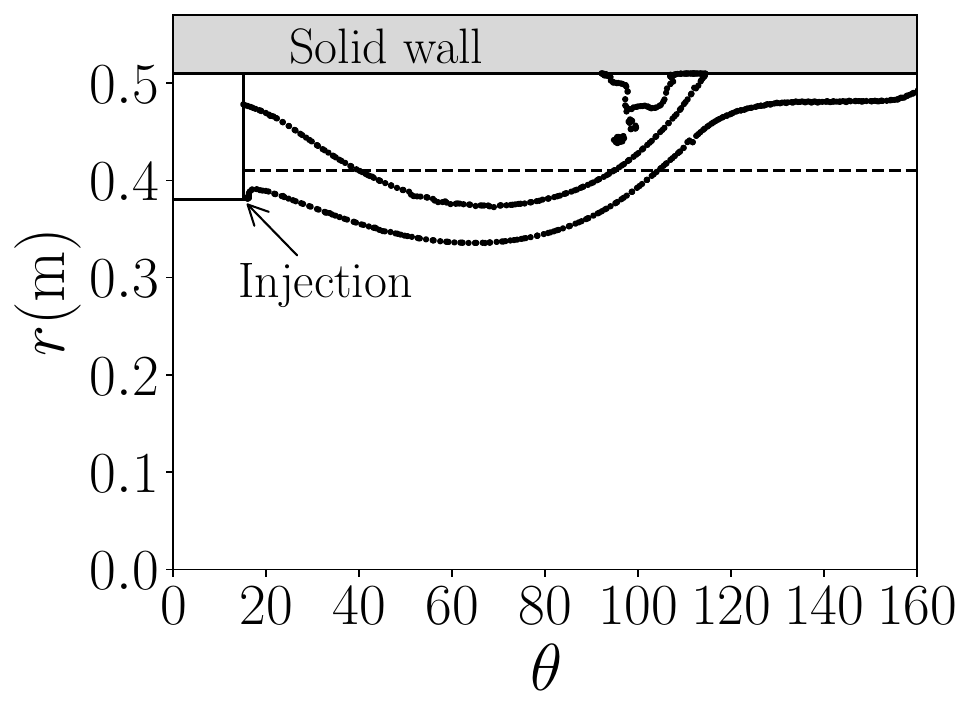}
\end{subfigure}
\begin{subfigure}{0.45\textwidth}
\caption{$Q = 10.5$ L/s, $I = 0$ A\label{fig:10.5L/s}}
\includegraphics[width =\textwidth]{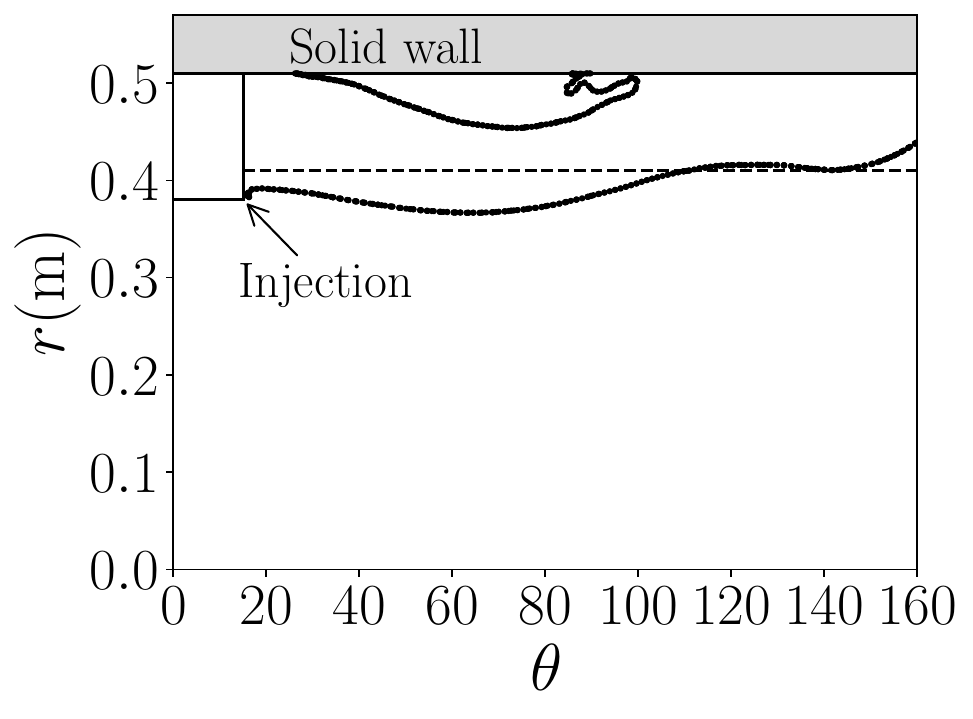}
\end{subfigure}
 \vspace{-0.5cm}
 \caption{\label{fig:freeSurface0A} Isoline positions of $\alpha = 0.5$, representing the interface between liquid ($\alpha = 1$) and gas ($\alpha = 0$), at $z = 0 \, \text{m}$ in polar coordinates $(r, \theta)$ for (a) 5.9 L/s and (b) 10.5 L/s. No current (and hence no Lorentz force) is applied in either case. As expected, the faster flow shows better adhesion to the solid wall. The dashed line at $r = 0.41 \, \text{m}$ indicates the target position of the liquid metal free surface.}
\vspace{-0.2cm}
\end{figure}
Experimental data \cite{EXPERIMENTALPAPER} are reported in Table \ref{tab:comparison}, in terms of free-surface velocity and jet thickness for a flow rate of \(Q = 9.3 \pm 0.2 \, \text{L/s}\). Numerical results are also reported, for a flow rate of \(Q = 9.2 \text{ L/s}\), showing excellent agreement with measurements.
\begin{table}[htb!]\centering
\caption{\label{tab:comparison} Experimental data for 400 rpm  as defined in \cite{EXPERIMENTALPAPER} and numerical results in the arc \(\theta < 36^\circ\), without injection of external current.}
\begin{tabular}{|c|c|c|}\hline
& Experimental data & Numerical results \\\hline
Flow rate (L/s) & $9.3\pm0.2$ & $9.2$
\\ 
Free-surface velocity (m/s)	& $1.7\pm0.2$ & 1.8 
\\ Thickness (m) & $0.09\pm0.01	$ & 0.09
\\\hline
\end{tabular}
\vspace{-0.2cm}
\end{table}



Figure \ref{fig:freeSurface0A} shows the position of the curves $\alpha=0.5$, representing the liquid-gas interface, in polar coordinates $(r, \theta)$ at \(z = 0 \, \text{m}\). Two flow rates (5.9 and $10.5 \text{ L/s}$) are reported, in the absence of external current. 
The intersection between the curve $\alpha = 0.5$ and the line $r = 0.51 \, \text{m}$ indicates the location of contact with the wall, signifying adhesion between the liquid phase and the solid wall. The inlet section into the chamber is located at $\theta=15^\circ$. In case (b), the injection section is fully filled, so $A_b = A_{\text{injection}}$, whereas in case (a), it is only partially filled, with $A_a \approx 0.75 A_{\text{injection}}$. Given that $Q_b \approx 1.8 Q_a$, we obtain $v_b \approx 1.4 v_a$, which leads to centrifugal forces being nearly twice as large in case (b), thus explaining the enhanced adhesion in that scenario.

 The thickness of the jet, being interpreted as the distance between the lower curve and the solid wall at \(r = 0.51 \, \text{m}\), increases with the flow rate, varying from a minimum of 2 cm for \(Q = 4.0 \, \text{L/s}\), to a maximum of 10 cm  for \(Q = 10.5 \, \text{L/s}\), with thickness of 3 cm and 8 cm for 5.9 L/s and 9.2 L/s respectively. This can be explained by the more intense relative effect of gravity for lower flow rates, as indicated by $Fr$ number in Table \ref{tab:nondimensional}.

\noindent
\subsubsection{{Controlled  flow - injection of external current}}
\begin{figure}[htb!]\vspace{-0.5cm}
\centering
\begin{subfigure}{0.45\textwidth}\caption{\label{fig:5.9L/s_currentScan}}
\includegraphics[width =\textwidth]{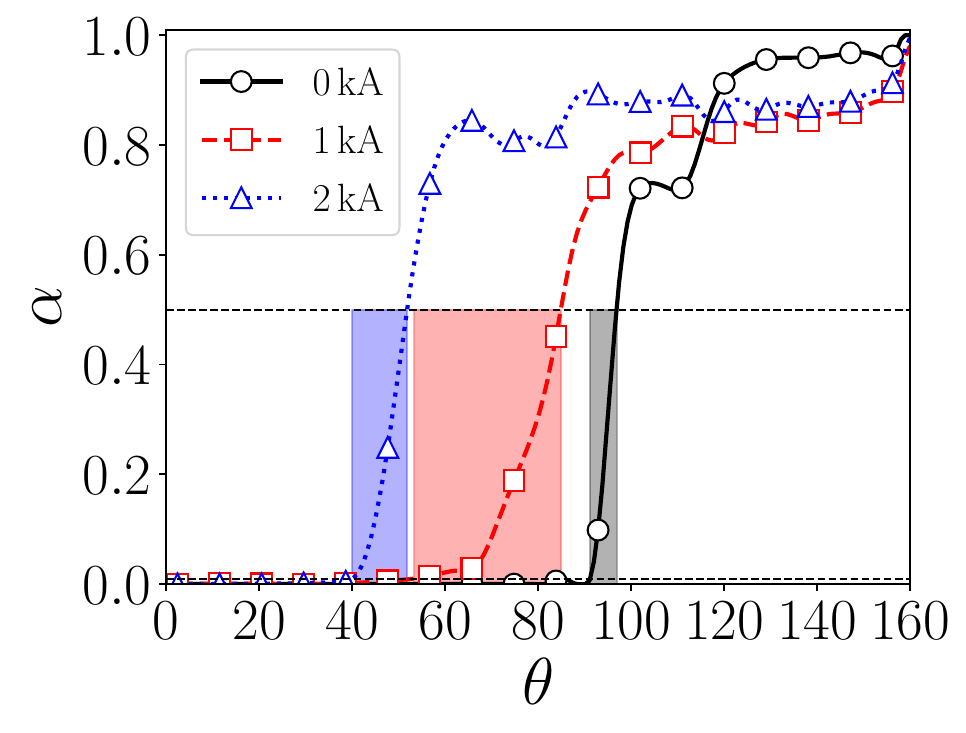}
\end{subfigure}
\begin{subfigure}{0.45\textwidth}\caption{\label{fig:exp}}
\includegraphics[width =\textwidth]{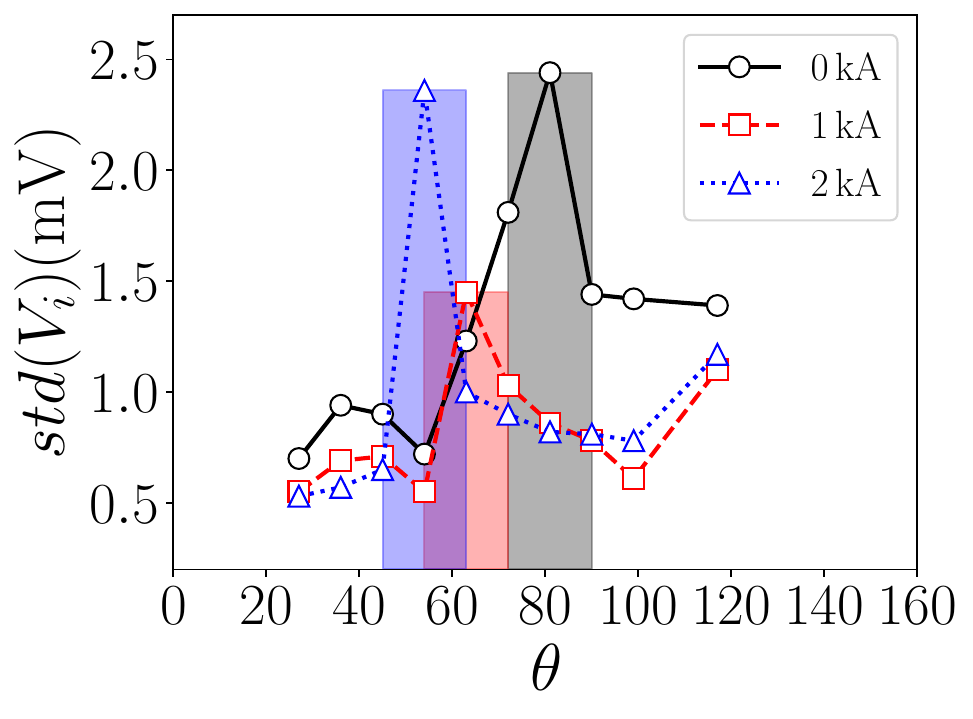}
\end{subfigure}
 \vspace{-0.5cm}
\caption{\label{fig:currentScan} Results of an electrical current scan (0 kA, 1 kA, and 2 kA). Increasing electrical currents - and thus electromagnetic forces - leads to better angular coverage and adhesion: (a) illustrated by the volume fraction $\alpha$ profiles for a flow rate of 5.9 L/s, and (b) experimentally measured by the maximum standard deviation of the electric potential at the electrodes for a flow rate of $6.2 \pm 0.2 \, \text{L/s}$ \cite{EXPERIMENTALPAPER}.}
\vspace{-0.5cm}
\end{figure}
\begin{figure}[htb!]\vspace{-0.5cm}
\centering
\begin{subfigure}{0.45\textwidth}
\caption{$Q = 5.9$ L/s, $I = 1$ kA\label{fig:5.95L/s1000A}}
\includegraphics[width =\textwidth]{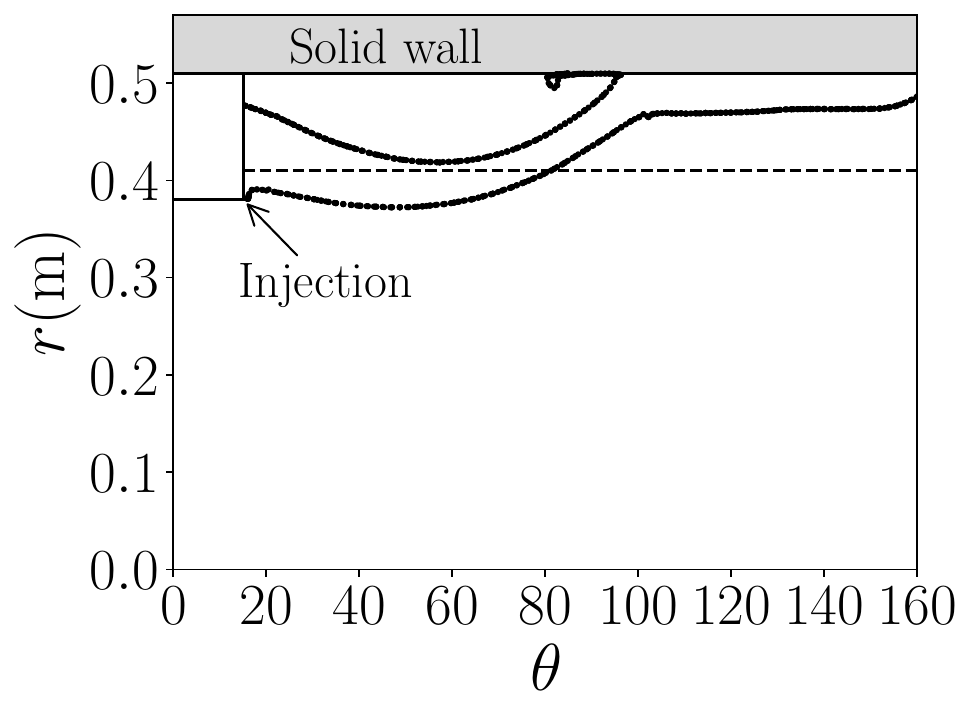}
\end{subfigure}
\begin{subfigure}{0.45\textwidth}
\caption{$Q = 10.5$ L/s, $I = 1$ kA\label{fig:10.5L/s1000A}}
\includegraphics[width =\textwidth]{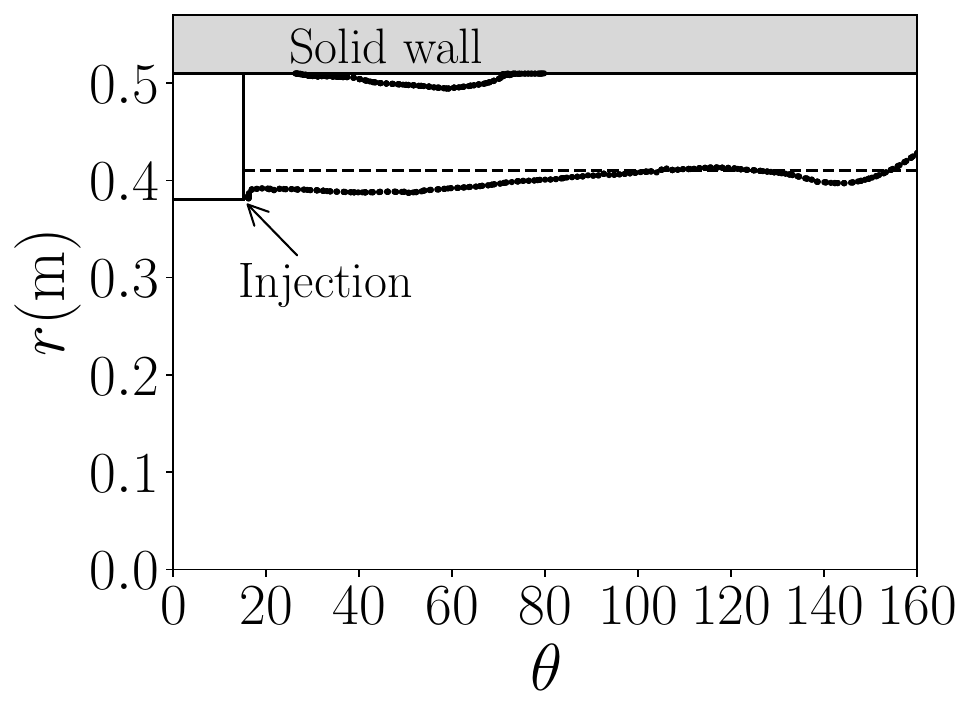}
\end{subfigure}
\begin{subfigure}{0.45\textwidth}
\caption{$Q = 5.9$ L/s, $I = 2$ kA\label{fig:5.95L/s2000A}}
\includegraphics[width =\textwidth]{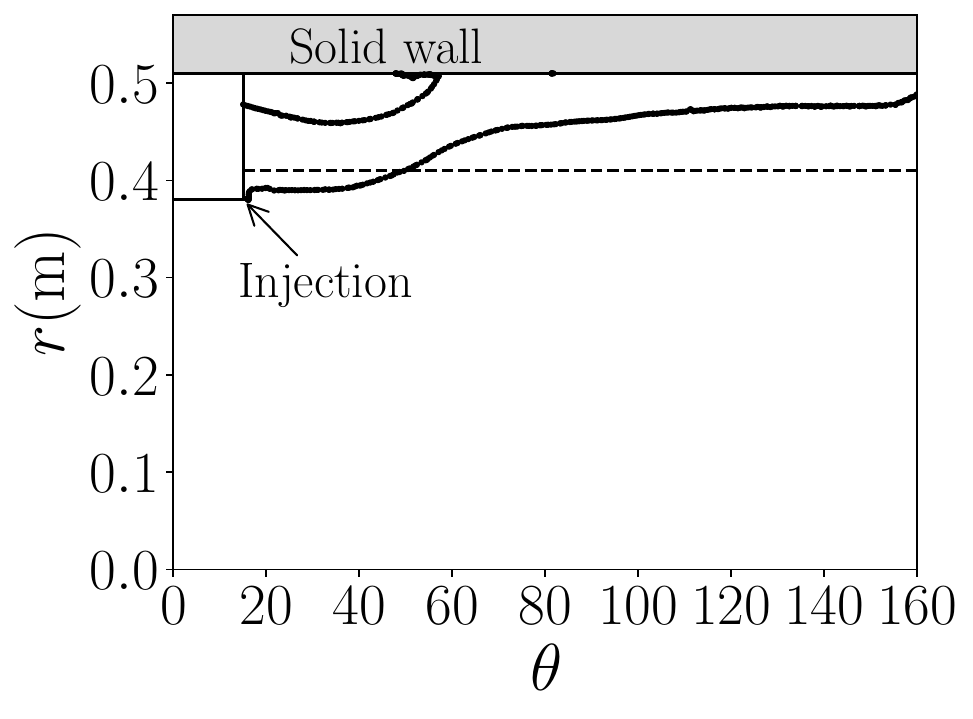}
\end{subfigure}
\begin{subfigure}{0.45\textwidth}
\caption{$Q = 10.5$ L/s, $I = 2$ kA\label{fig:10.5L/s2000A}}
\includegraphics[width =\textwidth]{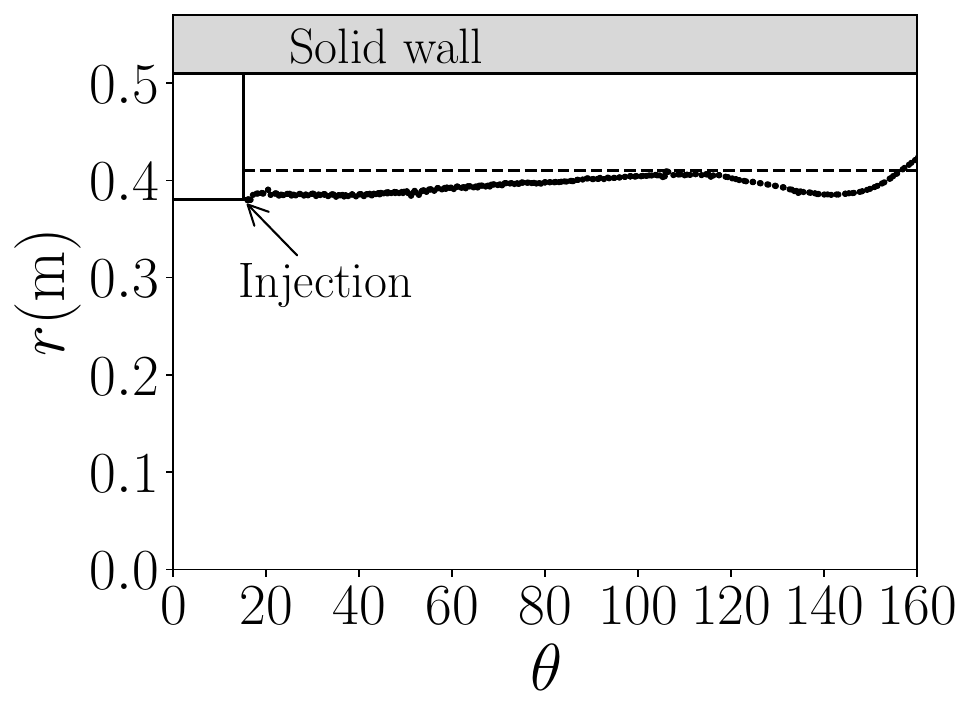}
\end{subfigure}\vspace{-0.5cm}
\caption{\label{fig:freeSurface1000A} Similar to Figure \ref{fig:freeSurface0A}, but with an applied current $I_{ext}=$ 1 kA and 2 kA. The resulting Lorentz force enhances adhesion to the wall compared to Figure \ref{fig:freeSurface0A}, as expected.
} \vspace{-0.5cm}
\end{figure}
In this section, results obtained with the injection of external current are presented.
Figure \ref{fig:5.9L/s_currentScan} shows the volume fraction \(\alpha\) plotted against the angular position \(\theta\) for a flow rate of 5.9 L/s. The base flow without current injection (0 kA) is compared with the profiles when external currents of 1 kA and 2 kA are applied.
The range of angular position $\theta$ corresponding to $0.01<\alpha <0.5$ is highlighted. The data have been averaged along $z$ direction and grouped with a bin width $\Delta \theta =5^\circ$.
In absence of external current, the landing point position, i.e. the location where the liquid metal jet enters in contact with the side walls, is in the range \(90^\circ < \theta < 100^\circ\).  When 1 kA are injected, the landing point shifts to \(60^\circ < \theta < 80^\circ\), and when 2 kA are injected, the landing point position shifts further up reaching \(40^\circ < \theta < 60^\circ\). 

 The values of standard deviation measured in the experimental campaign \cite{EXPERIMENTALPAPER} are reported in Figure \ref{fig:exp}, for \(Q = 6.2 \pm 0.2 \, \text{L/s}\). Experimentally, the landing point position is identified by the location of the electrode that measures the maximum standard deviation in the electrical potential signal. It has been observed that the landing point is at around $80^\circ$ without external current, shifts between $60^\circ$ and $70^\circ$ with the injection of 1 kA, and remains between $50^\circ$ and $60^\circ$ for 2 kA. These findings are in good agreement with the numerical results. 

Figure \ref{fig:freeSurface1000A} illustrates the radial coordinate \(r\) of the isoline \(\alpha = 0.5\) at the mid-plane (\(z = 0 \, \text{m}\)), plotted against the angular coordinate \(\theta\) for two flow rates (5.9 and 10.5 L/s) with an external current of 1 kA and 2 kA. From this plot series, we observe that the injection of external current shifts the landing point upwards and makes the jet thickness more uniform, especially at higher flow rates. In particular, we can observe that the case (d) shows a full adhesion of the liquid metal against the wall.


\section{Conclusions and future work}\label{sec:conclusion}
In this study, we developed a numerical model to simulate the behavior of liquid GaInSn within the vacuum chamber of the Skyfall 1b experiment at Renaissance Fusion. 
The numerical simulations are validated by experimental results. In particular, increasing the flow rate leads to a thicker jet and shifts the landing point. The landing point moves upwards with higher external current intensity, allowing better coverage control without adjusting the flow rate. This control is in good agreement with experimental data across a range of current injections (0 A to 2 kA) at a flow rate of 5.9 L/s, and has significant implications in fusion reactor applications, where consistent liquid coverage is essential for efficient cooling and shielding.
The model effectively captures key flow features, such as how the electromagnetic force impacts the liquid/gas interface and the current density dynamics. 

To further enhance the accuracy and robustness of the numerical model, future work should include refining the mesh resolution and performing sensitivity analyses across a broader range of parameters. For upcoming experiments involving hotter conditions, liquid tin, liquid lithium, and eventually liquid lithium-lithium hydride will replace the current working fluid. While the isothermal assumption may still hold in these experiments, it would no longer be valid in reactor environments, where significant temperature gradients between the inlet and outlet are expected.  
Regarding the assumption of a low magnetic Reynolds number ($R_m$), it remains valid for the 10 cm thick flows targeted in future experiments. However, in a reactor environment with liquid walls of approximately 40 cm thickness, this assumption will depend on the electrical properties of the working fluid. The lower conductivity of lithium-lithium hydride, as compared to pure liquid metals, ensures that $R_m$ remains below 1, maintaining the validity of this assumption for future fusion reactors.


\vspace{0.2cm}
\bibliography{biblio.bib}

\lastpageno
\end{document}